\documentclass[a4paper,12pt]{article}
\usepackage{graphicx}
\usepackage{hyperref}
\usepackage{latexsym}
\usepackage{color}
\usepackage{amsmath}
\usepackage{geometry}
\def\msbar{{\overline{\rm MS}}}

\title{Diquark mass differences from unquenched lattice QCD}
\author{Yujiang Bi$^1$\thanks{biyujiang@ihep.ac.cn}, Hao Cai$^1$\thanks{hcai@whu.edu.cn}, Ying Chen$^2$,
Ming Gong$^2$, \\
Zhaofeng Liu$^2$\thanks{liuzf@ihep.ac.cn}, Hao-Xue Qiao$^1$, and Yi-Bo Yang$^3$}

\date{}
\begin{document}
\maketitle

\begin{center}
$^1$School of Physics and Technology, Wuhan University, Wuhan 430072, China\\
$^2$Institute of High Energy Physics and Theoretical Physics Center for Science Facilities, Chinese Academy of Sciences, Beijing 100049, China\\
$^3$Department of Physics and Astronomy, University of Kentucky, Lexington, KY 40506, USA
\end{center}

\begin{abstract}
We calculate diquark correlation functions in the Landau
gauge on the lattice using overlap valence quarks and 2+1-flavor domain wall fermion configurations.
Quark masses are extracted from the scalar part of quark propagators in the Landau gauge.
Scalar diquark quark mass difference and axial vector scalar diquark mass difference are obtained 
for diquarks composed of two light quarks and of a strange and a light quark.
Light sea quark mass dependence of the results is examined. Two lattice spacings are used
to check the discretization effects. 
The coarse and fine lattices are of sizes $24^3\times64$
and $32^3\times64$ with inverse spacings $1/a=1.75(4) {\rm~GeV}$ and $2.33(5) {\rm~GeV}$, respectively.
\end{abstract}

\newpage
\section{Introduction}
Diquarks were introduced long time ago and are used in many phenomenology studies of strong interactions.
For example, diquarks are used to describe baryons with a quark-diquark picture for explaining the missing states~\cite{Anselmino:1992vg}.
Also they are used to explain the $\Delta I=1/2$ rule in weak nonleptonic decays~\cite{Dosch:1988hu,Neubert:1991zd}. 
A review on diquarks are given in Ref.~\cite{Anselmino:1992vg}.
Since the discovery of $X(3872)$ by the Belle Collaboration~\cite{Choi:2003ue}, experiments have observed many so called XYZ states~\cite{Agashe:2014kda}.
It is difficult to interpret these states as conventional heavy quarkonia. Molecules and tetraquarks are proposed to explain some of these quarkoniumlike states.
In the tetraquark scenario, a diquark and an antidiquark form a four quark state. 

Quantum Chromodynamics (QCD) is the theory describing the strong interaction among quarks and gluons,
from which the properties of diquark correlations (if any) should be understood.
At low energies, QCD has to be solved by nonperturbative methods since the strong
coupling constant becomes so strong that perturbative calculations break down.
In this work we study diquarks starting from the QCD action by using lattice QCD simulations.
Diquarks have been studied on the lattice
from various approaches~\cite{Hess:1998sd,Alexandrou:2006cq,Babich:2007ah,Orginos:2005vr,Fodor:2005qx,DeGrand:2007vu,Green:2010vc}
to deal with the fact that they are not color singlets.
Diquark masses were also calculated in QCD sum rules, see, for example, Refs.~\cite{Kleiv:2013dta,Wang:2011ab}. 
And the stability of diquarks were discussed as well~\cite{Tang:2012np}.

The scalar diquark is supposed to be the state with the strongest correlation. 
The mass difference between
diquarks with various quantum numbers can reflect the relative size of the correlation.
In Ref.~\cite{Jaffe:2004ph},
scalar diquark and quark mass difference as well as the axial vector and scalar diquark mass difference are estimated from baryon spectroscopy.
On the lattice, diquark mass and mass differences can be studied in a fixed gauge. 
So far, the masses and mass differences are calculated mostly in the quenched approximation on the lattice. Here
we calculate them by using $2+1$ flavor domain wall fermion configurations. For the valence quark, we use overlap fermions.
Previous lattice calculations focused on diquarks composed of the light up and down quarks. In this work we consider diquarks
composed of a strange and a light quark as well as of two light quarks. 

Diquark correlations are induced by spin dependent interactions. Thus they become weaker as the
masses of quark increase. We can look into the mass dependence of diquark correlations by varying the current quark masses on the lattice. 

Our results of the diquark mass difference and diquark quark mass difference are summaried in Tab.~\ref{tab:summary}.
In general they agree with the estimations in Ref.~\cite{Jaffe:2004ph}.
The exception is the scalar diquark and strange quark mass difference for diquarks composed of a strange and a light quark. 
Our result of this difference is smaller than the estimation in Ref.~\cite{Jaffe:2004ph}.

The paper is organized as follows. In section~\ref{sec:setup}, we present the details of our calculation.
The results and discussion are given in section~\ref{sec:results}. 
Finally we summarize in section~\ref{sec:summary}.

\section{Correlation functions and lattice setup}
\label{sec:setup}
The two quarks in the scalar diquark ($J^P=0^+$) in the color antitriplet representation are attractive to each other as favored by perturbative
one gluon exchange~\cite{DeGrand:1975cf} and by instanton interactions~\cite{'tHooft:1976fv,Schafer:1996wv}. 
It is often called a ``good" diquark. The next favored diquark is the axial vector diquark
($J^P=1^+$) and is called a ``bad" diquark. The other diquarks with $J^P=0^-,1^-$ are supposed to be even energetic and therefore their masses are
higher. The diquarks in the color sextet representation have much larger color electrostatic field energy and are not favored 
by various models~\cite{Jaffe:2004ph}.

Therefore we focus on the color antitriplet diquarks and calculate their masses on the lattice.
The interpolating operators used in this work for diquarks with various quantum numbers are 
given in Tab.~\ref{tab:currents}, where $C$ is the charge conjugation operator. In the operators $q_1$ and $q_2$ are $u$ and $d(s)$ respectively.
\begin{table}
\caption{Interpolating fields and correlation functions of diquarks. A trace is performed in color space.}
\begin{center}
\begin{tabular}{cll}
\hline\hline
$J^P$ (diquark) & Operators & Correlators \\
\hline
$0^+$(good) &  $J^5_c=\epsilon_{abc}[q_1^aC\gamma_5q_2^b]$ &
$\sum_{\vec x}\langle\Omega|TJ^5_c(x)\bar{J}^5_c(0)|\Omega\rangle$ \\
$0^+$(good) &  $J^{05}_c=\epsilon_{abc}[q_1^aC\gamma_0\gamma_5q_2^b]$ &
$\sum_{\vec x}\langle\Omega|TJ^{05}_c(x)\bar{J}^{05}_c(0)|\Omega\rangle$ \\
$1^+$(bad) &  $J^i_c=\epsilon_{abc}[q_1^aC\gamma_iq_2^b]$ &
$\frac{1}{3}\sum_i\sum_{\vec x}\langle\Omega|TJ^i_c(x)\bar{J}^i_c(0)|\Omega\rangle$ \\
$1^+$(bad) &  $J_c=\epsilon_{abc}[q_1^a q_1^b]$ & $\sum_{\vec x}\langle\Omega|TJ_c(x)\bar{J}_c(0)|\Omega\rangle$ \\
$0^-$ &  $J^I_c=\epsilon_{abc}[q_1^aCq_2^b]$ &
$\sum_{\vec x}\langle\Omega|TJ^I_c(x)\bar{J}^I_c(0)|\Omega\rangle$ \\
$0^-$ &  $J^0_c=\epsilon_{abc}[q_1^aC\gamma_0 q_2^b]$ &
$\sum_{\vec x}\langle\Omega|TJ^0_c(x)\bar{J}^0_c(0)|\Omega\rangle$ \\
$1^-$ &  $J^{i5}_c=\epsilon_{abc}[q_1^aC\gamma_i\gamma_5q_2^b]$ &
$\frac{1}{3}\sum_i\sum_{\vec x}\langle\Omega|TJ^{i5}_c(x)\bar{J}^{i5}_c(0)|\Omega\rangle$ \\
\hline\hline
\end{tabular}
\label{tab:currents}
\end{center}
\end{table}
In calculating the two point functions in Tab.~\ref{tab:currents}, we take a trace in color space and project to zero momentum.
Diquarks are not color singlet. Their two point correlation functions have to be computed in a fixed gauge. We use the Landau gauge.

Alternatively, one can combine a diquark with an infinitely heavy quark (i.e. a Polyakov line) to get a color singlet state and
calculate its correlation function. This is gauge invariant, however it may have path dependence~\cite{Hess:1998sd}.

We use the RBC-UKQCD configurations generated with 2+1-flavor domain wall fermions~\cite{Aoki:2010dy}. The parameters of the ensembles used
in this work are given in Table~\ref{tab:confs}.
\begin{table}
\begin{center}
\caption{Parameters of configurations with 2+1 flavor dynamical domain wall fermions (RBC-UKQCD).
The residual masses are from Ref.~\cite{Aoki:2010dy}.
The lattice spacings are from Ref.~\cite{Yang:2014sea}.}
\begin{tabular}{cccccc}
\hline\hline
$1/a$(GeV) & label & $am_{sea}$ & volume & $N_{src}\times N_{conf}$ & $am_{res}$\\
\hline
1.75(4) & c005  & 0.005/0.04 & $24^3\times64$ & $8\times92$ & 0.003152(43) \\
& c02 & 0.02/0.04 & $24^3\times64$ & $8\times99$ & \\
\hline
2.33(5) & f004 & 0.004/0.03 & $32^3\times64$ & $1\times50$ & 0.0006664(76) \\
\hline\hline
\end{tabular}
\label{tab:confs}
\end{center}
\end{table}
On the coarse lattice, two different light sea quark masses are used to check the sea quark
mass dependence of our results. To examine finite lattice spacing effects, we use one ensemble on a fine lattice. To improve the signal,
on each configuration on the coarse lattice we compute eight point source quark propagators. The sources are evenly located on
eight time slices. On each time slice the source position is randomly chosen from one configuration to another to reduce data correlations.
For the vector and axial vector diquarks, we average over the three directions ($i=1,2,3$) to increase statistics. Also we average the
data in the forward and backward time directions.

For the valence quark, we use overlap fermions.
The massless overlap operator~\cite{Neuberger:1997fp} is defined as
\begin{equation}
D_{ov}  (\rho) =   1 + \gamma_5 \varepsilon (\gamma_5 D_{\rm w}(\rho)).
\end{equation}
Here $\varepsilon$ is the matrix sign function and $D_{\rm w}(\rho)$ is the usual Wilson fermion operator, 
except with a negative mass parameter $- \rho = 1/2\kappa -4$ in which $\kappa_c < \kappa < 0.25$. 
We use $\kappa = 0.2$ in our calculation that corresponds to $\rho = 1.5$. The massive overlap Dirac operator is defined as
\begin{eqnarray}
D_m &=& \rho D_{ov} (\rho) + m\, (1 - \frac{D_{ov} (\rho)}{2}) \nonumber\\
       &=& \rho + \frac{m}{2} + (\rho - \frac{m}{2})\, \gamma_5\, \varepsilon (\gamma_5 D_w(\rho)).
\end{eqnarray}
To accommodate the $SU(3)$ chiral transformation, it is usually convenient to use the chirally regulated field
$\hat{\psi} = (1 - \frac{1}{2} D_{ov}) \psi$ in lieu of $\psi$ in the interpolation field and operators.
This is equivalent to leaving the unmodified operators and instead adopting the effective quark propagator
\begin{equation}
G \equiv D_{eff}^{-1} \equiv (1 - \frac{D_{ov}}{2}) D^{-1}_m = \frac{1}{D_c + m},
\end{equation}
where $D_c = \frac{\rho D_{ov}}{1 - D_{ov}/2}$ is chiral, i.e. $\{\gamma_5, D_c\}=0$~\cite{Chiu:1998gp}.

The overlap valence quark masses in lattice units are given in Tab.~\ref{tab:24_32}. 
Using the quark mass renormalization constants from Ref.~\cite{Liu:2013yxz} and the lattice spacings, 
one finds the corresponding $\msbar$ quark masses at 2 GeV are from about 20 MeV to 1 GeV.
\begin{table}
\begin{center}
\caption{Overlap valence quark masses in lattice units used in this work.}
\begin{tabular}{cccccccc}
\hline\hline
$24^3\times64$  & 0.01350  & 0.02430 & 0.04890 & 0.06700 & 0.15000 & 0.33000 & 0.67000 \\
\hline
$32^3\times64$ &  0.00677 &  0.01290 & 0.02400 & 0.04700 & 0.18000 & 0.28000 & 0.50000 \\ 
\hline\hline
\end{tabular}
\label{tab:24_32}
\end{center}
\end{table}
The bare quark masses $am_q=0.067$ and $0.047$ on the coarse and fine lattices correspond to the physical strange quark mass respectively
within our uncertainty~\cite{Yang:2014sea}. Our largest quark mass is less than but close to the charm quark mass.

\section{Results and discussion}
\label{sec:results}
In this section, we give the numerical results on all three ensembles given in Tab.~\ref{tab:confs}. 
The statistical errors of our results are from bootstraps with 500 samples.
On the coarse lattice, we have two ensembles with different
light sea quark masses. Thus we only check the sea quark mass dependence in the results but do not try to extrapolate to the chiral limit of the
sea quark. The results on the fine lattice, ensemble f004, are used to check the discretization effects but
have relatively large uncertainty due to the limited statistics. 

\subsection{Pion, nucleon and Delta masses}
For the convenience of chiral extrapolation later, we first give the pion masses from our data.
We also compute the two point correlation functions for the nucleon and $\Delta^{++}$ baryon, from which we extract their masses at our pion masses.
We use the usual interpolating operators $\bar u\gamma_5 d$, $\epsilon_{abc}[u^{Ta}(C\gamma_5)d^b] u^c$ and $\epsilon_{abc}[u^{Ta}(C\gamma_\mu)u^b]u^c$
for these hadrons respectively. The correlation functions are projected to zero 3-momentum and to positive parity (for the baryons).
At large $t$, the ground state dominates and the hadron mass is obtained from a single exponential fit to the correlator.

The numerical values of these hadron masses are given in Tabs.~\ref{tab:hadrons},~\ref{tab:hadrons_c02},~\ref{tab:hadrons_f004}
for ensemble c005, c02 and f004 respectively. On ensemble f004, the signal to noise ratio for the correlators of the Delta baryon is
too bad for us to obtain its mass, especially at light quark masses.
\begin{table}
\begin{center}
\caption{Pion, nucleon and Delta masses from ensemble c005. $1/a=1.75(4)$ is used to convert to physical units.}
\begin{tabular}{cccccc}
\hline\hline
$am_{q}$       & $am_{ps}$ & $m_{ps}/$GeV & $am_N$ & $am_{\Delta^{++}}$ & $a(m_{\Delta^{++}}-m_N)$ \\
\hline
0.01350   & 0.17911(46) & 0.3134(72) &   0.677(17) & 0.818(41) & 0.141(39) \\
0.02430   & 0.23670(43) & 0.4142(95) & 0.7218(61) &  0.863(20) & 0.141(19) \\
0.04890   & 0.33520(55) & 0.587(13) & 0.8043(64) &  0.924(18) & 0.120(18) \\
0.06700   & 0.39173(46) & 0.686(16) & 0.8643(47) & 0.973(11) & 0.109(12) \\
0.15000     & 0.60552(35) & 1.060(24) & 1.1247(42) & 1.2049(66) & 0.0802(72) \\
0.33000    & 0.97666(36) & 1.709(39) &1.6398(22) & 1.6991(27) & 0.0593(26) \\
\hline\hline
\end{tabular}
\label{tab:hadrons}
\end{center}
\end{table}
At leading order in heavy baryon chiral perturbation theory~\cite{Gasser:1987rb},
the nucleon mass is given by 
\begin{equation}
M_N=M_0-4c_1m_\pi^2-\frac{3g_A^2}{32\pi f_\pi^2}m_\pi^3.
\label{eq:baryon_3terms}
\end{equation}
Here we take the experiment values 1.267 for $g_A$ and 92.4 MeV for $f_\pi$ to fit our nucleon mass as a function of the pion mass.
Note by doing this, we have ignored the effects from the mixed action setup.
Half of our pion masses are larger than 600 MeV. Thus we do not expect the pion mass dependence of all the baryon masses can be well
described by formulae from chiral perturbation theory. In using Eq.(\ref{eq:baryon_3terms}) to do the fit, we only include the 
data points with pion mass less than 600 MeV.

\begin{figure}
\begin{center}
\includegraphics[height=2.5in,width=0.49\textwidth]{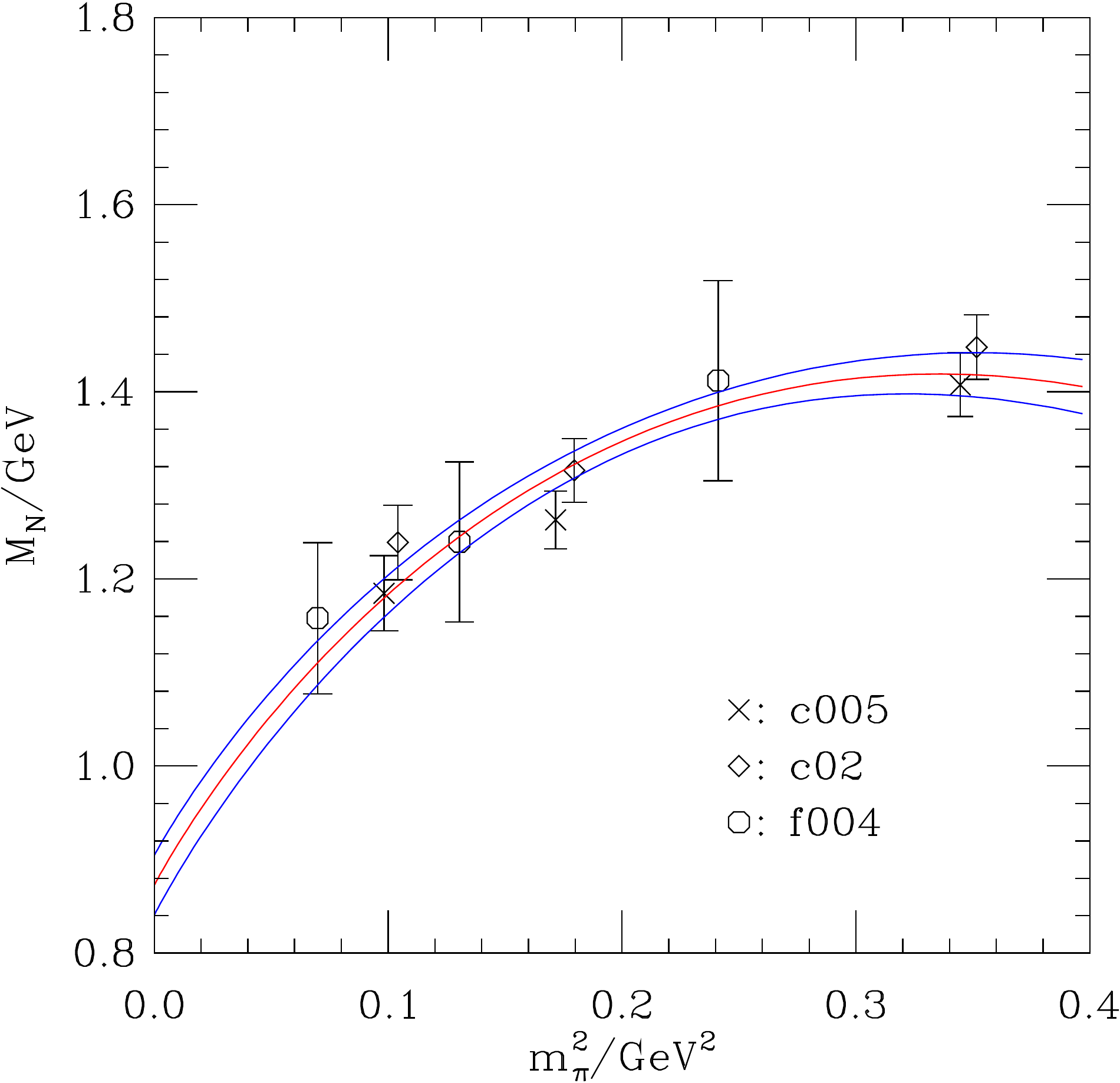}
\includegraphics[height=2.5in,width=0.49\textwidth]{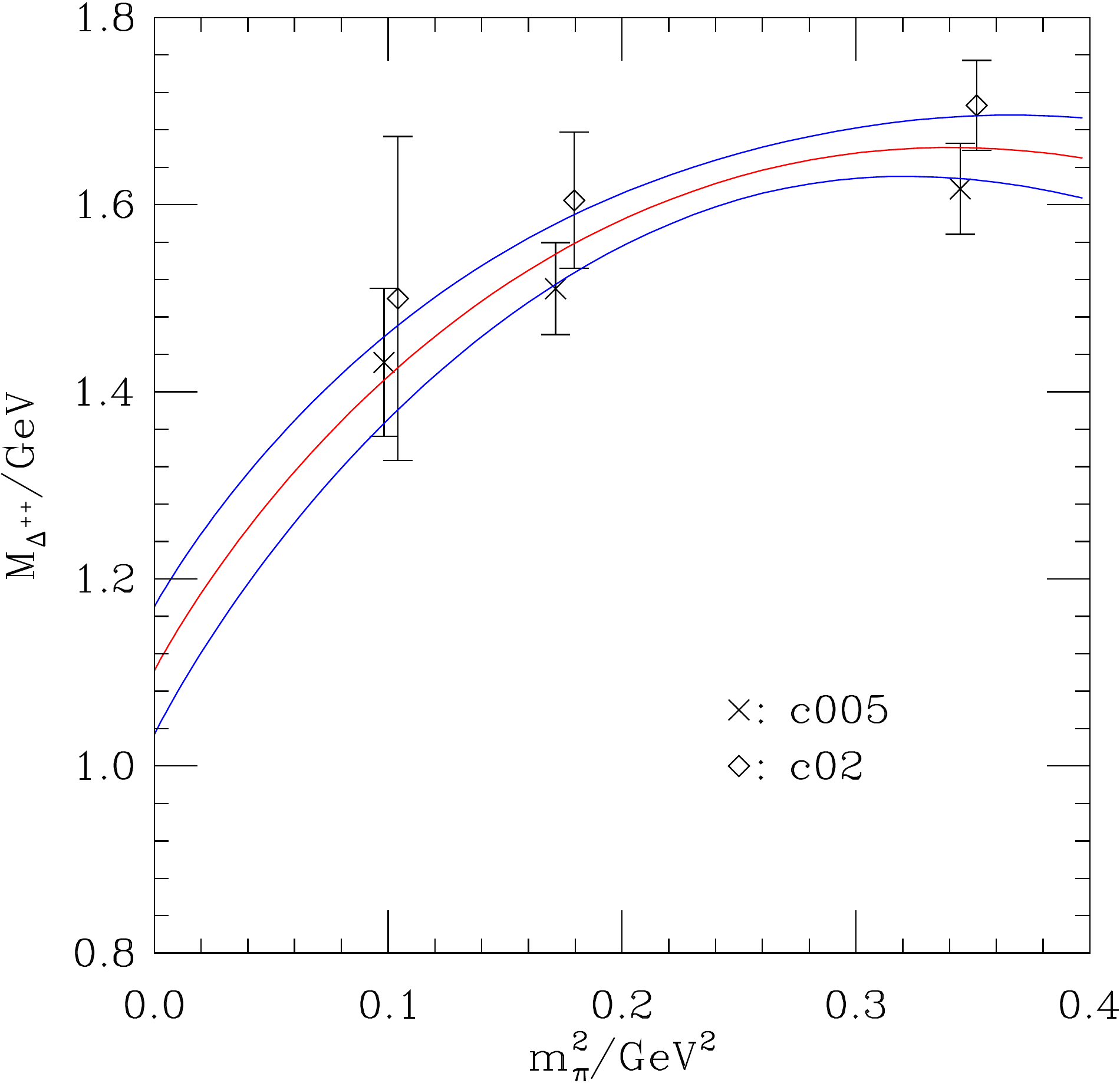}
\end{center}
\caption{Left: The chiral fit to the nucleon mass using data from all three ensembles at pion mass less than 600 MeV.
Right: The chiral fit to the mass of $\Delta^{++}$ using data from ensembles c005 and c02 at pion masses less than 600 MeV.}
\label{fig:Mn_Mpi2}
\end{figure}

In the left graph of Fig.~\ref{fig:Mn_Mpi2}, the nucleon masses from all three ensembles are plotted against the pion mass squared.
Here and in the rest of the paper, we have taken into account the uncertainties of the lattice spacings
in converting our results into physical units.
We do not see big sea quark mass dependence or discretization effects.
The two parameter ($M_0$ and $c_1$) fit is shown by the red curve and has a $\chi^2$ per degree
of freedom (dof) equal to 0.73. At the physical pion mass 140 MeV, the fit gives $m_N=953(30)$ MeV, 
which agrees with the experimental value 940 MeV. 
A three parameter function $M_N=M_0+c_1 m_\pi^2+c_2 m_\pi^3$ can also fit the same data with a good $\chi^2$. It gives
a nucleon mass (1.09(9) GeV) at the physical point. This is similar to the chiral fits in Refs.~\cite{Gong:2013vja,Liu:2014jua}.

For the pion mass dependence of the Delta baryon, the leading one loop result from heavy baryon chiral perturbation theory has a
same form as in Eq.(\ref{eq:baryon_3terms}) if the SU(6) relation $H_A=9g_A/5$ is used for the Delta baryon axial coupling $H_A$.
The fit of Eq.(\ref{eq:baryon_3terms}) to the data points at pion mass less than 600 MeV on ensembles c005 and c02 is shown in
the right graph of Fig.~\ref{fig:Mn_Mpi2}. The $\chi^2/$dof of the fit is 0.73 and 
we get $m_{\Delta}=1.183(64)$ GeV at the physical pion mass. This should be compared with the experiment value $1232$ MeV with a
width 117 MeV.

In the last column of Tab.~\ref{tab:hadrons} and Tab.~\ref{tab:hadrons_c02}, 
we give the mass difference between the Delta and the nucleon in lattice units.
From Eq.(\ref{eq:baryon_3terms}), we see this mass difference is a linear function of the pion mass squared at leading order.
Extrapolating the lowest three data points to the physical pion mass using a linear function of $(am_\pi)^2$, one gets 
$a(m_{\Delta^{++}}-m_N)=0.155(32)$ or $m_{\Delta^{++}}-m_N=272(56)$ MeV for ensemble c005.
\begin{table}
\begin{center}
\caption{Pion, nucleon and Delta masses from ensemble c02. $1/a=1.75(4)$ is used to convert to physical units.}
\begin{tabular}{cccccc}
\hline\hline
$am_{q}$       & $am_{ps}$ & $m_{ps}/$GeV & $am_N$ & $am_{\Delta^{++}}$ & $a(m_{\Delta^{++}}-m_N)$ \\
\hline
0.01350   & 0.18440(63) & 0.3227(75) &   0.708(16) & 0.857(97) & 0.149(95) \\
0.02430   & 0.24213(53) & 0.4237(97) & 0.7520(91)  &  0.917(36) & 0.165(35) \\
0.04890   & 0.33899(56) & 0.593(14)  & 0.8272(56)  &  0.975(16) & 0.148(20) \\
0.06700   & 0.39692(55) & 0.695(16)  & 0.8936(41)  & 1.0313(94) & 0.138(12) \\
0.15000   & 0.60939(47) & 1.066(24)  & 1.1622(28)  & 1.2584(51) & 0.0962(57) \\
0.33000   & 0.98015(30) & 1.715(39)  & 1.6594(23)  & 1.7196(33) & 0.0602(40) \\
\hline\hline
\end{tabular}
\label{tab:hadrons_c02}
\end{center}
\end{table}
Similarly for ensemble c02, we get $a(m_{\Delta^{++}}-m_N)=0.174(62)$ or $m_{\Delta^{++}}-m_N=304(108)$ MeV.
Both values agree with the experiment value $\sim292$ MeV but have large uncertainties.
\begin{table}
\begin{center}
\caption{Pion and nucleon masses from ensemble f004. $1/a=2.33(5)$ is used to convert to physical units.}
\begin{tabular}{cccc}
\hline\hline
$am_{q}$       & $am_{ps}$ & $m_{ps}/$GeV & $am_N$ \\
\hline
0.00677   & 0.1134(35) & 0.2642(99) &   0.497(33)  \\
0.01290   & 0.1550(26) & 0.3612(98) & 0.532(35)   \\
0.02400   & 0.2106(15) & 0.491(11)  & 0.606(44)  \\
0.04700   & 0.2970(10) & 0.692(15)  & 0.672(12)  \\
0.18000   & 0.6321(10) & 1.473(32)  & 1.0848(72)  \\
0.28000   & 0.8394(11) & 1.956(42)  & 1.3851(44)  \\
\hline\hline
\end{tabular}
\label{tab:hadrons_f004}
\end{center}
\end{table}

\subsection{Quark correlation functions}
\label{sec:quark_mass}
\begin{figure}
\begin{center}
\includegraphics[height=2.5in,width=0.49\textwidth]{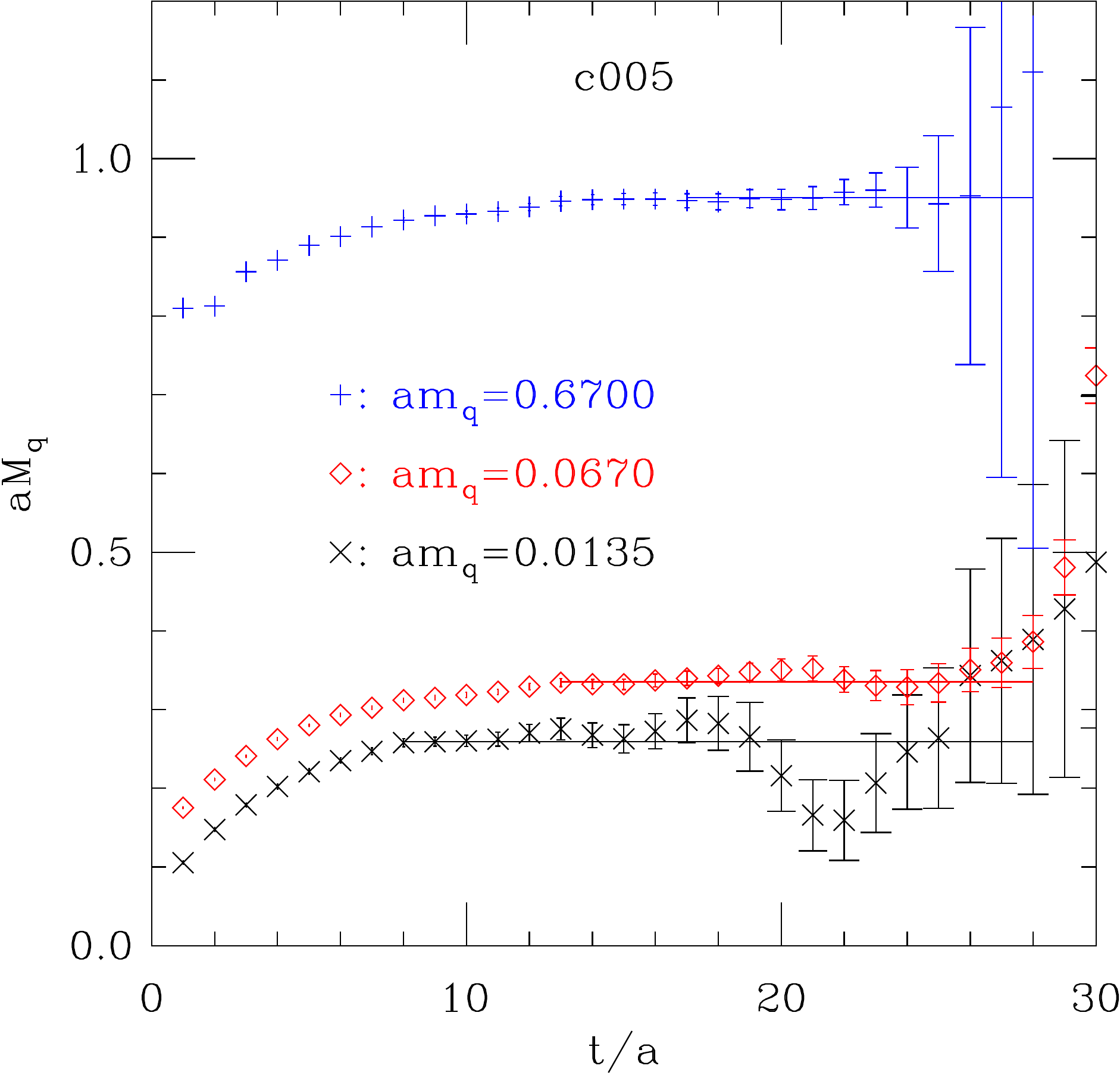}
\end{center}
\caption{The effective quark mass $\ln(C_q(t)/C_q(t+1))$
at various bare valence quark masses for ensemble c005 with sea quark masses $m_l/m_s=0.005/0.04$. 
The straight lines mark the results of the quark mass from single exponential fits to the correlators $C_q(t)$. 
The fitting ranges of $t$ are also indicated by the lines.}
\label{fig:mq_eff_c005}
\end{figure}

The diquark-quark mass difference is a measure of the strength of the diquark correlation. This difference for the good diquark
is estimated from hadron spectrum in Ref.~\cite{Jaffe:2004ph}. We obtain the quark mass from the scalar part of the quark propagator
$S(x,0)$ in Landau gauge
\begin{equation}
C_q(t)=\sum_{\vec x}S_{ii}^{aa}(x,0),
\end{equation}
where the color index $a$ and spin index $i$ are summed over. A single exponential fit (actually a hyperbolic sine function because of the
boundary condition in the time direction) taking into
account data correlations to $C_q(t)$ at large $t$, for example $t\in[13,28]$,
 gives the quark
mass $aM_q$ at each bare valence quark mass. Examples of the effective mass $\ln(C_q(t)/C_q(t+1))$ and the results from the exponential fits are 
shown in Fig.~\ref{fig:mq_eff_c005}.

\begin{figure}
\begin{center}
\includegraphics[height=2.5in,width=0.49\textwidth]{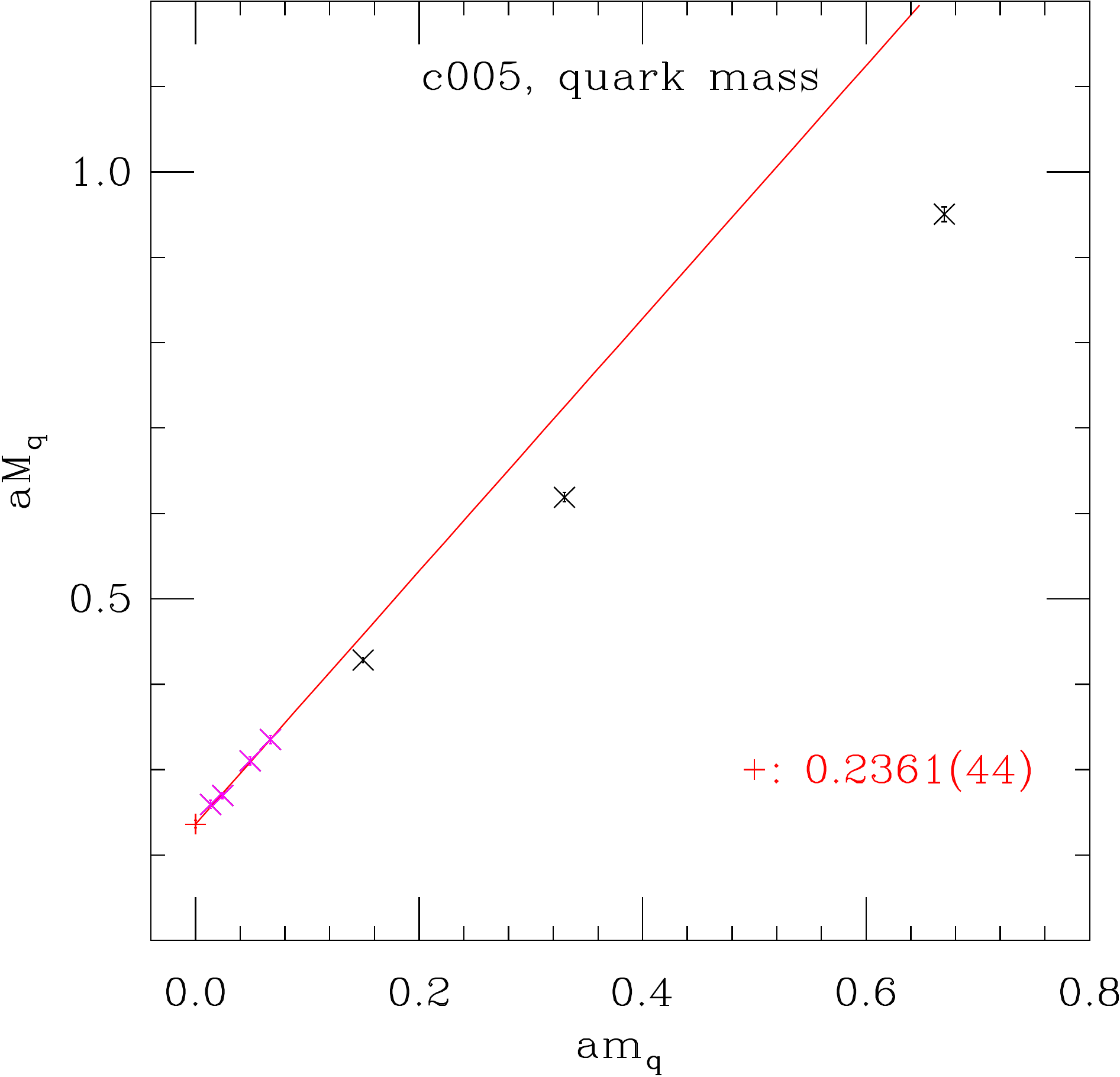}
\includegraphics[height=2.5in,width=0.49\textwidth]{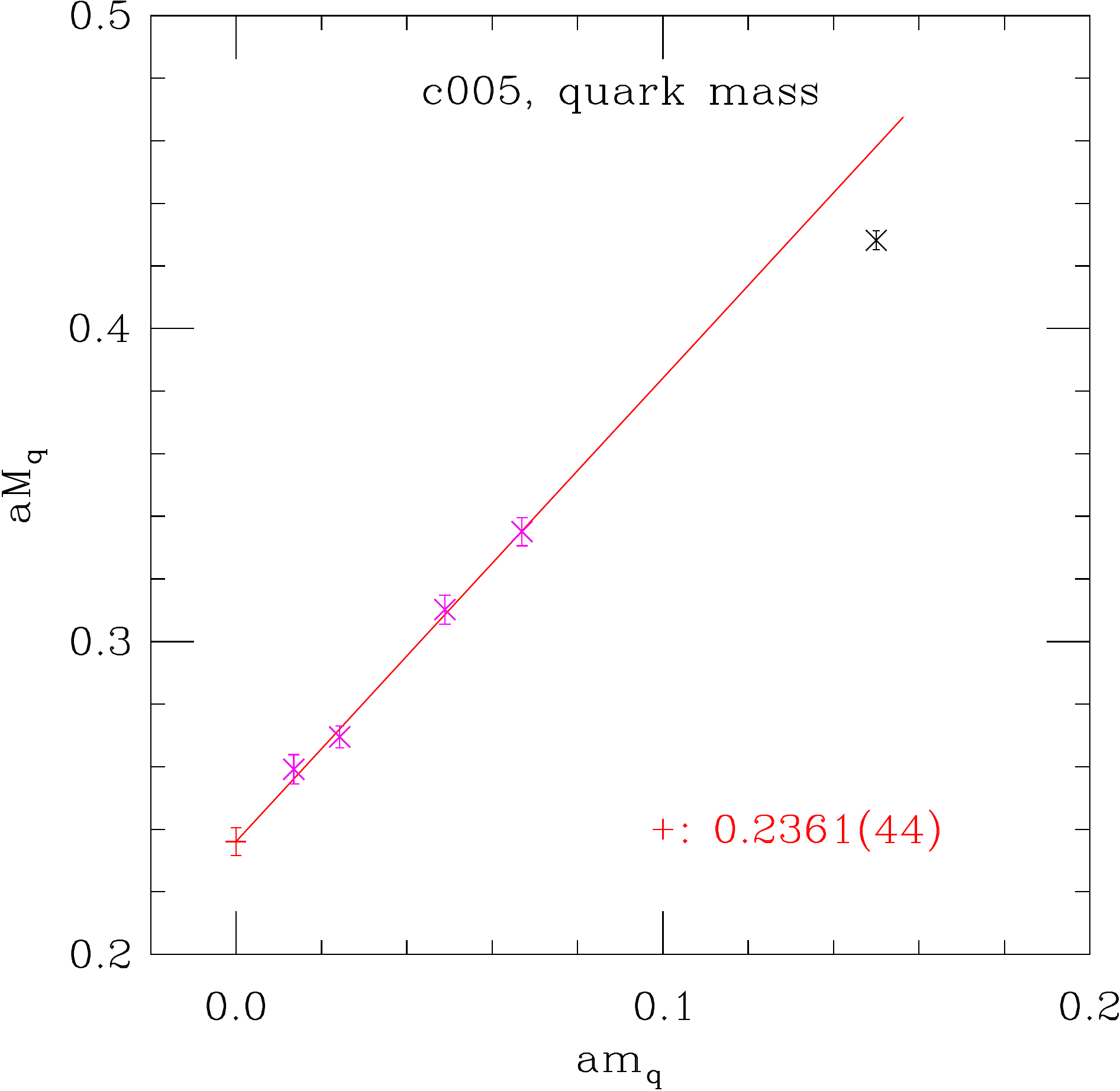}
\end{center}
\caption{Quark masses $aM_q$ against the bare valence quark mass $am_q$ for ensemble c005 with sea quark masses $m_l/m_s=0.005/0.04$. 
The straight line is a linear fit to the lightest four quark masses. The graph on the right is a zoom in of the left.}
\label{fig:amq_c005}
\end{figure}

The quark masses $aM_q$ are collected in Tab.~\ref{tab:quark_mass} for all three ensembles.
\begin{table}
\begin{center}
\caption{Quark masses for various valence quark masses on all three ensembles. The first line is a linear extrapolation in $am_q$ to the chiral
limit with the lowest four quark masses.}
\begin{tabular}{ccc|cc}
\hline\hline
$am_q$(coarse)      & $aM_q$(c005) & $aM_q$(c02) & $am_q$(fine) & $aM_q$(f004) \\
\hline
0.0     & 0.2361(44) & 0.2813(86) & 0.0 & 0.1832(99) \\
0.01350  &  0.2592(47) & 0.298(12) & 0.00677 & 0.196(14) \\
0.02430  &  0.2695(35) &  0.3034(73) & 0.01290 & 0.199(10) \\
0.04890   & 0.3102(46) &  0.3258(43) & 0.02400 & 0.2144(89) \\
0.06700  &  0.3351(45) & 0.3443(37) & 0.04700  & 0.2466(81) \\
0.15000 & 0.4282(30) &  0.4362(32) & 0.18000 & 0.3893(79) \\
0.33000 & 0.6190(57) &  0.6264(58) & 0.28000 & 0.5006(92) \\
0.67000  & 0.9504(85) & 0.9559(88) & 0.50000 & 0.721(12) \\
\hline\hline
\end{tabular}
\label{tab:quark_mass}
\end{center}
\end{table}
They are plotted against the bare quark mass for ensemble c005 in Fig.~\ref{fig:amq_c005}.
With a linear fit to the lightest four quark masses (the corresponding current quark masses are not heavier than the physical strange quark mass),
one finds in the chiral limit $aM_q=0.2361(44)$ or $M_q=413(12)$ MeV by using $1/a=1.75(4)$ GeV. 
The fitting is also shown in Fig.~\ref{fig:amq_c005} by the red line. 

\begin{figure}
\begin{center}
\includegraphics[height=2.5in,width=0.49\textwidth]{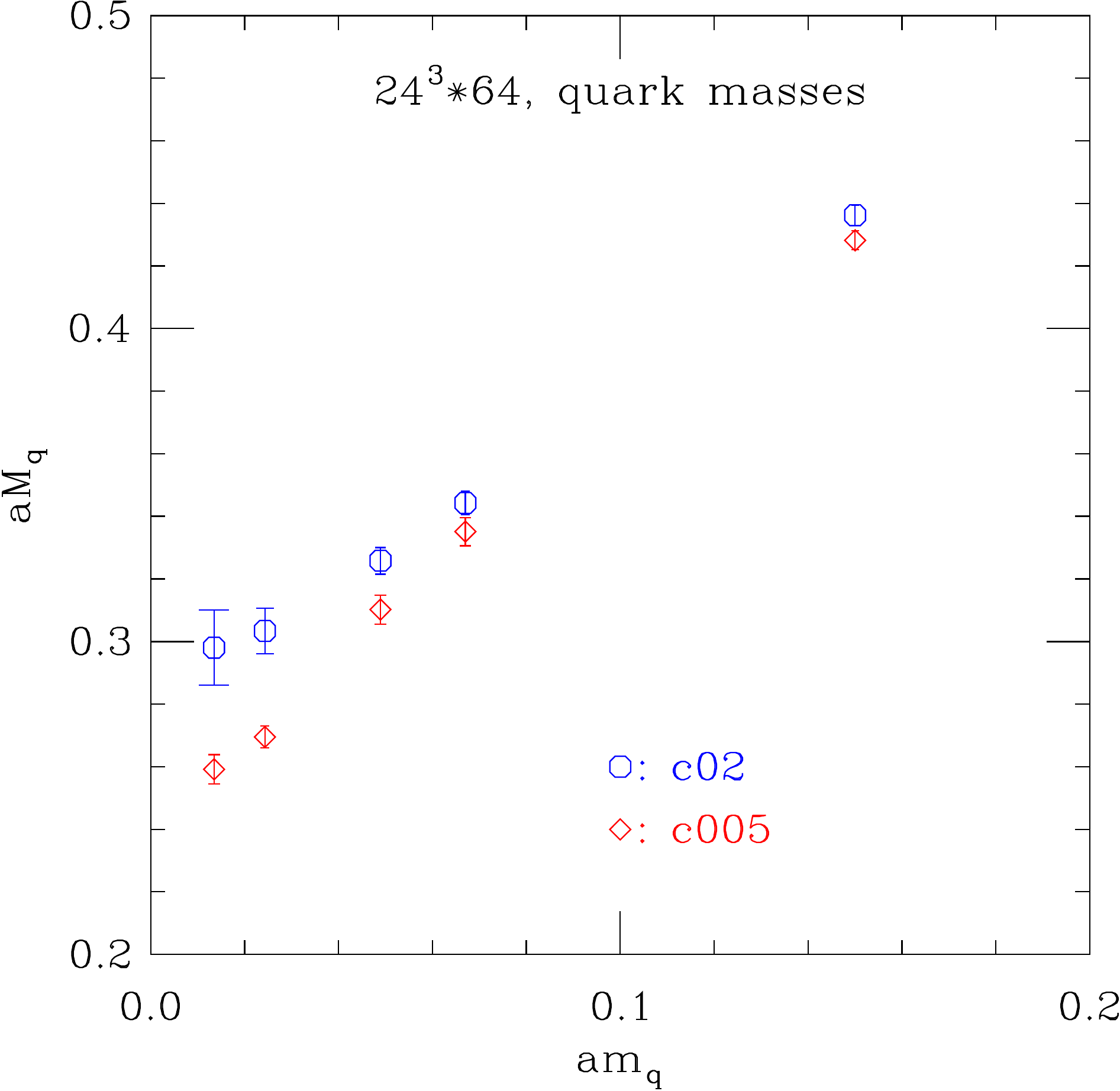}
\end{center}
\caption{Quark masses from the scalar part of quark propagators against the bare valence quark masses on the two coarse lattices with different
sea quark masses.}
\label{fig:amq_coarse}
\end{figure}

Similarly on the ensemble c02, we obtain $aM_q=0.2813(86)$ or $M_q=492(19)$ MeV in the chiral limit by a linear extrapolation
with the lowest four quark masses. The two ensembles c02 and c005 have different
sea quark masses. To see the sea quark mass dependence more clearly, we plot $aM_q$ from the two ensembles together in Fig.~\ref{fig:amq_coarse}.
We see that there is a clear sea quark mass dependence in our results 
when the valence quark mass is less than the physical strange quark mass ($am_q\le0.067$).
The quark mass $aM_q$ from the scalar part of the quark propagator in Landau gauge decreases as the sea quark mass decreases.

The results from ensemble f004 are also given in Tab.~\ref{tab:quark_mass}. In the chiral limit of the valence quark mass, we get
$aM_q=0.1832(99)$ or $M_q=427(25)$ MeV by using $1/a=2.33(5)$ GeV. If using the lowest five data points to do the linear extrapolation,
then we find $aM_q=0.1889(58)$ or $M_q=440(16)$ MeV.
The light sea quark mass for ensemble f004 is lighter than but close to that for c005. Since the number $M_q=427(25)$ MeV agrees with
the result 413(12) MeV from c005, we do not see apparent discretization effects in $M_q$ with our statistical uncertainty.

Although the quark mass $M_q$ here is in principle gauge dependent, it was shown this dependence may be small in covariant gauges including
the Landau gauge~\cite{Bernard:1989ih}. If we average the quark masses $M_q$ from c005 and f004 
using the inverse of their squared error
as the weight, then we obtain $M_q=416(11)$ MeV. Here the inverse of the square of the final statistical uncertainty
is equal to the sum of the inverse squared error (the weight). Note our sea quark mass is still larger than the physical one.
$M_q$ is in general consistent with constituent 
quark masses (350-400 MeV) used in various models.

For the strange quark, its bare valence quark mass is roughly 0.067 and 0.047 on the coarse and fine lattice respectively~\cite{Yang:2014sea}.
The corresponding quark mass $M_s$ from the scalar part of the quark propagator is 586(16), 603(15) and 575(23) MeV
on the three ensembles c005, c02 and f004 respectively (see Tab.~\ref{tab:quark_mass}). Unlike the light quark mass $M_q$, the sea quark mass dependence in
$M_s$ is much smaller.

\subsection{Diquark masses and mass differences}
\subsubsection{Diquarks composed of two light quarks}
\begin{figure}
\begin{center}
\includegraphics[height=2.5in,width=0.49\textwidth]{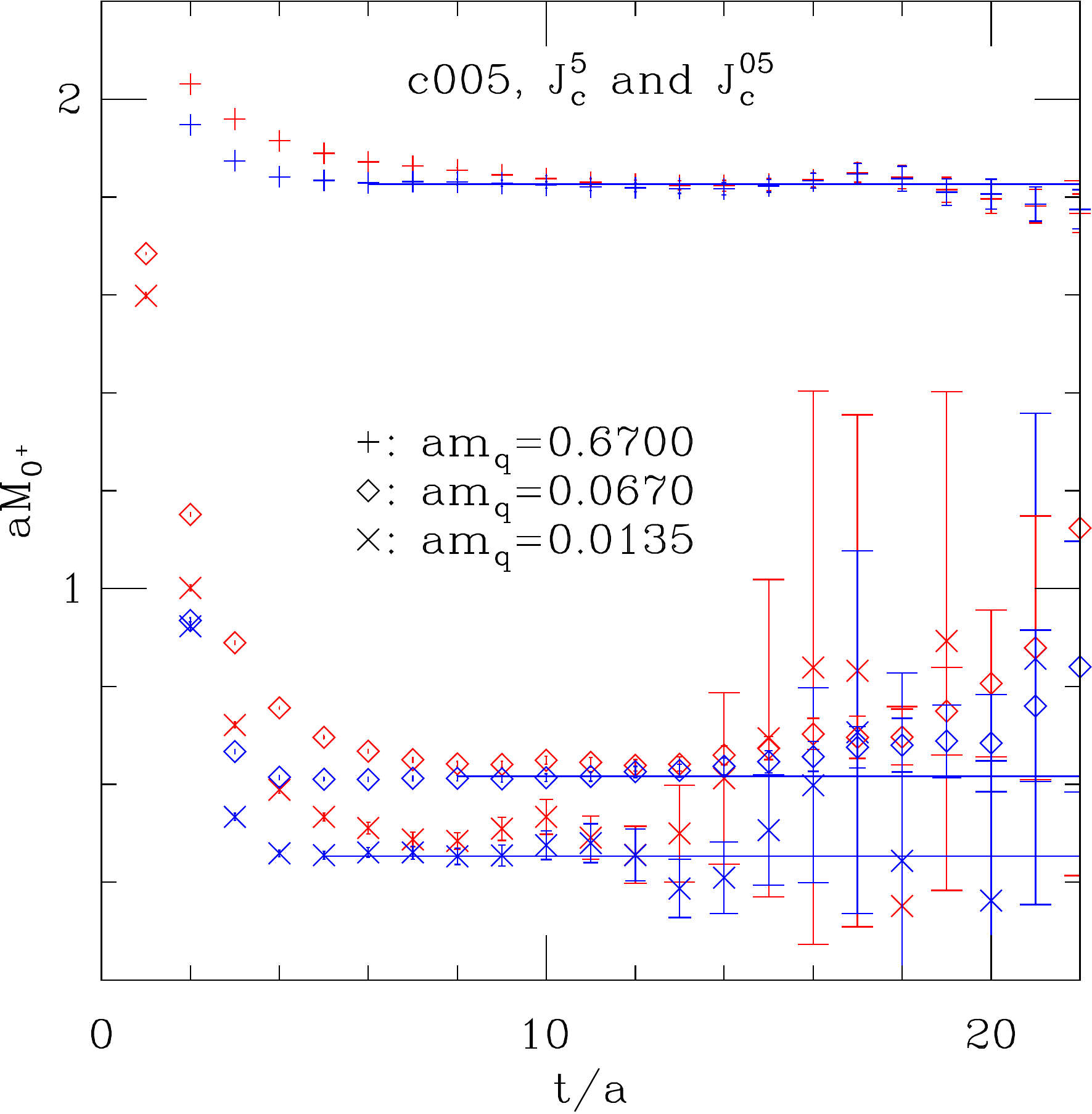}
\end{center}
\caption{Effective scalar diquark masses at various valence quark masses on ensemble c005 with sea quark masses $m_l/m_s=0.005/0.04$. 
The red symbols are from the correlators from $J^{5}_c$. The blue ones from $J^{05}_c$.
The straight lines mark the results of the scalar diquark mass from single exponential fits to the correlation functions using $J^{05}_c$. 
The fitting ranges of $t$ are indicated by the length of the lines.}
\label{fig:scalar_c005}
\end{figure}

We start with diquarks composed of the up and down quarks, i.e. $q_1=u$ and $q_2=d$ in Tab.~\ref{tab:currents}. 
The two point functions of diquarks exhibit an exponential decay behaviour similar to
that of ordinary hadron correlation functions. This can be clearly seen in the effective mass plot shown by Fig.~\ref{fig:scalar_c005} 
from the correlation functions using $J^5_c$ and $J^{05}_c$. 
Both currents $J^5_c$ and $J^{05}_c$ can give us the scalar diquark mass. The effective masses from the two currents go to a same plateau at large $t$
as we see in the graph. But the plateau for $J^{05}_c$ shows up earlier than that for $J^5_c$. Therefore we fit the correlators from $J^{05}_c$
to determine the scalar diquark mass.
In Fig.~\ref{fig:scalar_c005} the lines indicate the results from single exponential fittings to the correlators at different quark masses.

The numerical results of the scalar diquark mass are 
given in Tab.~\ref{tab:diquark_mass} for c005. 
With a linear chiral extrapolation, the scalar diquark mass from $J^{05}_c$ is $725(20)$ MeV by using $1/a=1.75(4)$ GeV.
\begin{table}
\begin{center}
\caption{Diquark masses and mass difference for various valence quark masses on ensemble c005. The first line is a linear extrapolation in $am_q$ to the chiral
limit with the lowest four data points.}
\begin{tabular}{cccccc}
\hline\hline
$am_q$       & $aM_{0^+}$($J^{05}_c$) & $aM_{1^+}$($J^i_c$) & $a(M_{1^+}-M_{0^+})$ & $aM_{0^-}$($J_c^I$) & $aM_{1^-}$ \\
\hline
0.0      &  0.4142(63) &  0.584(21) & 0.166(22) & - & - \\
0.01350   & 0.4534(70) & 0.611(29) & 0.158(31) & - & - \\
0.02430   & 0.4875(52)  & 0.635(18) & 0.148(19) & 0.796(52) & - \\
0.04890   & 0.5692(37)  & 0.694(10) & 0.1248(98) & 0.862(23) & 0.987(53) \\
0.06700   & 0.6166(48) & 0.7300(85) & 0.1134(93) & 0.904(18) & 1.003(41) \\
0.15000  &  0.8293(70)  &  0.8907(68) & 0.0614(89) & 1.056(29) & 1.140(24) \\
0.33000  &  1.1830(30) & 1.2334(55) & 0.0504(45) & 1.378(17) & 1.454(21) \\
0.67000  & 1.8265(39) & 1.8604(68) & 0.0339(62) & 1.976(12) & 2.025(16) \\
\hline\hline
\end{tabular}
\label{tab:diquark_mass}
\end{center}
\end{table}
Then the scalar diquark and quark mass difference from ensemble c005 
is $725(20)-413(12)=312(23)$ MeV. Here the final uncertainty is from a simple error propagation.
This number is in good agreement with the estimation $\sim310$ MeV in Ref.~\cite{Jaffe:2004ph}.

The results from the ensemble c02 are obtained similarly and are collected in Tab.~\ref{tab:diquark_mass_c02}. 
\begin{table}
\begin{center}
\caption{Diquark masses and mass difference for various valence quark masses on ensemble c02. 
The first line is a linear extrapolation in $am_q$ to the chiral
limit with the lowest four data points.}
\begin{tabular}{cccccc}
\hline\hline
$am_q$       & $aM_{0^+}$($J^{05}_c$) & $aM_{1^+}$($J^i_c$) & $a(M_{1^+}-M_{0^+})$ & $aM_{0^-}$($J_c^I$) & $aM_{1^-}$ \\
\hline
0.0      &  0.4555(91) &  0.644(16) &  0.185(20) & - & - \\
0.01350   & 0.491(10) & 0.662(21) & 0.171(26) & - & -\\
0.02430   & 0.5256(80)  & 0.687(13) & 0.161(16) & 0.900(57) & - \\
0.04890   & 0.5998(75)  & 0.727(11) &  0.127(12) & 0.950(22) & 0.956(79) \\
0.06700   & 0.6453(63) & 0.7574(85) & 0.1121(95) & 0.984(16) & 1.011(64) \\
0.15000  &  0.8521(76)  &  0.9145(58) &  0.0624(87) & 1.104(13) & 1.165(33) \\
0.33000  &  1.2060(56) & 1.2459(56) &  0.0399(53) & 1.400(19) & 1.441(25) \\
0.67000  & 1.836(11) &  1.8588(94) & 0.0228(77) & 1.969(14) & 2.053(16) \\
\hline\hline
\end{tabular}
\label{tab:diquark_mass_c02}
\end{center}
\end{table}
In the chiral limit, the scalar diquark mass from ensemble c02 is $797(24)$ MeV by using $1/a=1.75(4)$ GeV. This value is a little different from
725(20) MeV for ensemble c005. Thus there seems to be some sea quark mass dependence in the absolute value of the scalar diquark mass.
The scalar diquark and quark mass difference is $797(24)-492(19)=305(31)$ MeV, which agrees with the result 312(23) MeV from ensemble c005.
Therefore we do not see sea quark mass dependence in the diquark quark mass difference.

The results on the fine lattice f004 are given in Tab.~\ref{tab:diquark_mass_f004}.
\begin{table}
\begin{center}
\caption{Diquark masses and mass difference for various valence quark masses on ensemble f004. 
The first line is a linear extrapolation in $am_q$ to the chiral
limit with the lowest four data points.}
\begin{tabular}{cccccc}
\hline\hline
$am_q$       & $aM_{0^+}$($J^{05}_c$) & $aM_{1^+}$($J^i_c$) & $a(M_{1^+}-M_{0^+})$ & $aM_{0^-}$($J_c^I$) & $aM_{1^-}$ \\
\hline
0.0      &  0.296(19) & 0.425(24)  &  0.128(30) & - & - \\
0.00677   & 0.318(28) & 0.438(46) & 0.120(54) & - & -\\
0.01290   & 0.340(22)  & 0.460(24) & 0.120(32) & - & - \\
0.02400   & 0.379(15)  & 0.487(18) &  0.108(23) & - & - \\
0.04700   & 0.457(10) &  0.547(15) & 0.090(18) & 0.61(10)   & 0.565(17) \\
0.18000  &  0.7879(84) & 0.835(11)  &  0.047(13) & 0.957(20)& 0.829(19) \\
0.28000  &  0.9977(84) & 1.031(10) &  0.033(11) & 1.136(26) & 1.017(13) \\
0.50000  & 1.4299(90)  & 1.447(14)  & 0.017(15) & 1.592(16) & 1.454(15) \\
\hline\hline
\end{tabular}
\label{tab:diquark_mass_f004}
\end{center}
\end{table}
The scalar diquark mass in the chiral limit is $aM_{0^+}=0.296(19)$ or $M_{0^+}=690(47)$ MeV.
With a relatively large error, it is in agreement with the number 725(20) MeV from ensemble c005, which indicates the 
finite lattice spacing effect is smaller than our statistical uncertainty. The mass difference between the scalar
diquark and the light quark is $690(47)-427(25)=263(53)$ MeV. It just agrees with the results 312(23) MeV and 305(31) MeV
from c005 and c02 respectively.

We average the scalar diquark and quark mass difference from the three ensembles using the inverse of their squared error
as the weight. The sum of the inverse squared error gives the inverse of the final uncertainty squared.
In this way, we obtain $M_{0^+}-M_q=304(17)$ MeV, where the error is statistical (including the uncertainties of lattice spacing).

The mass of the bad diquark can be extracted from the correlators using $J^i_c$ or $J_c$. The fitting results by using a single exponential
from the two currents are in agreement. However the signal of the correlator from $J^i_c$ is better. Therefore we collect the bad diquark masses
from this current in Tab.~\ref{tab:diquark_mass} for c005. A linear extrapolation to the chiral limit with the lowest four data points gives 
$aM_{1^+}=0.584(21)$ or $M_{1^+}=1022(44)$ MeV. 

\begin{figure}
\begin{center}
\includegraphics[height=2.5in,width=0.49\textwidth]{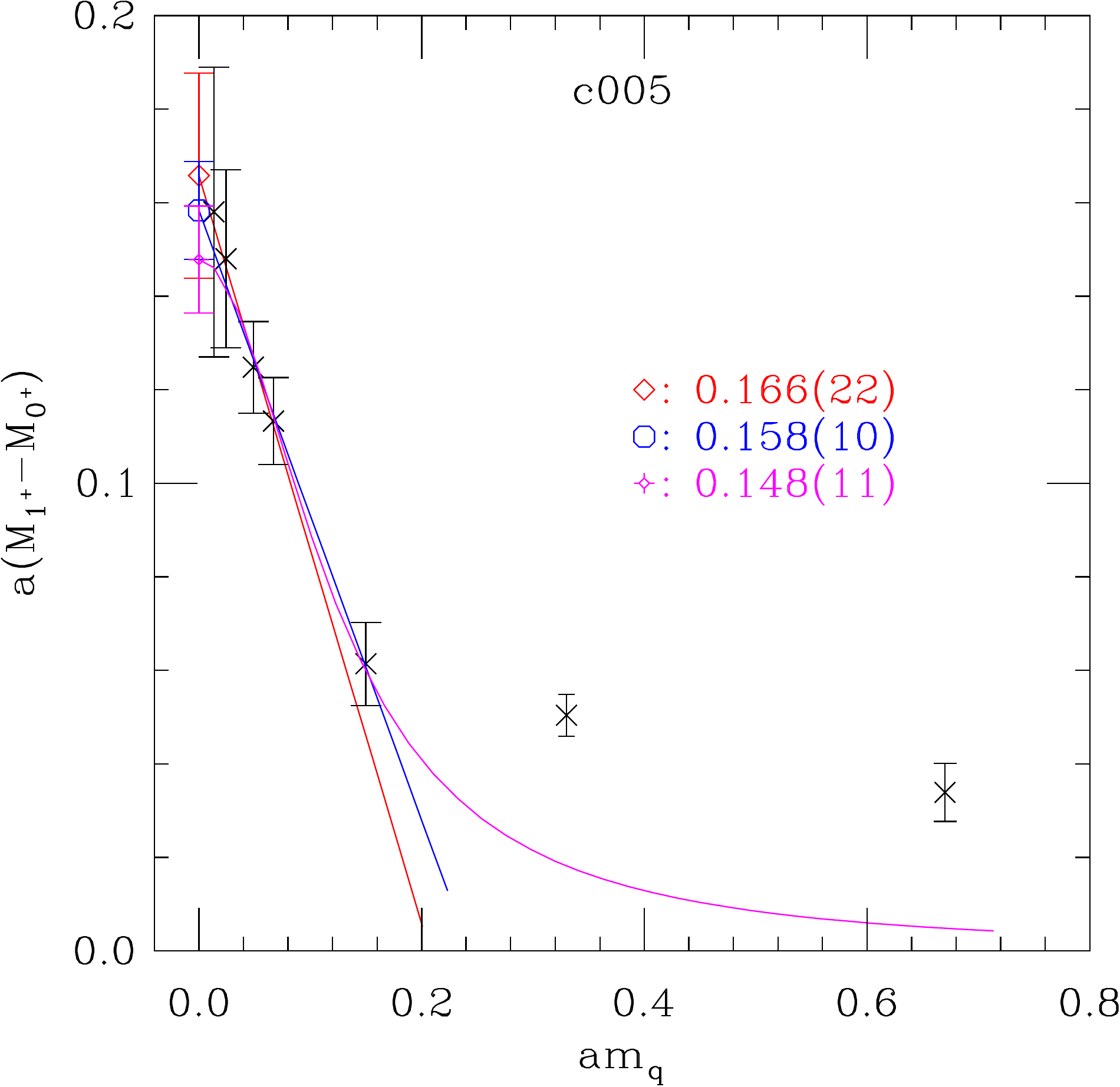}
\end{center}
\caption{The mass difference between bad and good diquarks as a function of the valence quark mass on ensemble c005.
The diquarks are composed of two degenerate light quarks.
The red straight line is a linear extrapolation in $am_q$ to the chiral limit using the lowest four data points. The blue
straight line uses the lowest five data points. The magenta curve is the fit using Eq.(\ref{eq:ans1}) to the lowest five points.}
\label{fig:diff_ud_diquark_c005}
\end{figure}

The bad and good diquark mass difference (in lattice units) is plotted against the valence quark mass in Fig.~\ref{fig:diff_ud_diquark_c005}. 
Here the uncertainties are from bootstrap analysis. As we can see, the difference decreases as the quark mass increases.
The simplest chiral extrapolation of the mass difference is a straight line fit. 
Using the lowest four data points, one gets $a(M_{1^+}-M_{0^+})=0.166(22)$ or
$M_{1^+}-M_{0^+}=291(39)$ MeV. 
If we use the lowest five data points for the linear chiral extrapolation, then we get $a(M_{1^+}-M_{0^+})=0.158(10)$, which agrees with
the result by using the lowest four data points. 

We also tried a
ansatz similar to the one used in Ref.~\cite{Alexandrou:2006cq}. Diquark correlations come from spin dependent forces. 
The bad and good diquark mass difference $\Delta m$ is expected to scale like $1/(m_{q_1}m_{q_2})$ at large quark mass~\cite{Jaffe:2004ph}. 
In the chiral limit, the mass difference goes to a constant. Thus one can try the following ansatz for diquarks composed of two degenerate quarks
\begin{equation}
\Delta m=\frac{a_1}{1+a_2 m_q^2},
\label{eq:ans1}
\end{equation}
where $a_1$ and $a_2$ are two fitting parameters. We find that this ansatz can not fit all seven data points with an acceptable $\chi^2/$dof.
If we limit to the lowest five data points, we can get a $\chi^2/$dof$<1$ and find $a(M_{1^+}-M_{0^+})=0.148(11)$. The fit
is shown by the curve in Fig.~\ref{fig:diff_ud_diquark_c005}.

\begin{figure}
\begin{center}
\includegraphics[height=2.5in,width=0.49\textwidth]{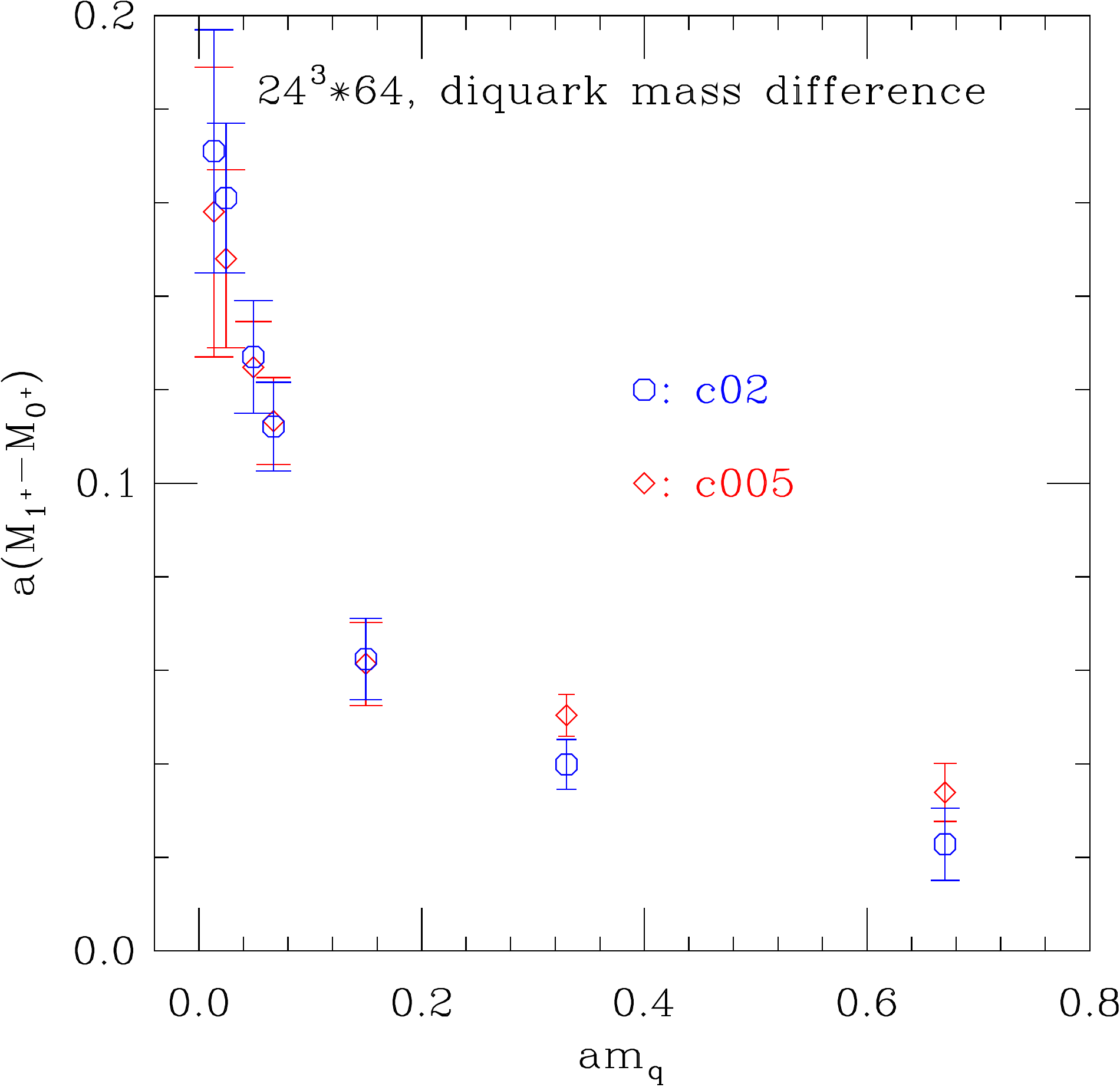}
\includegraphics[height=2.5in,width=0.49\textwidth]{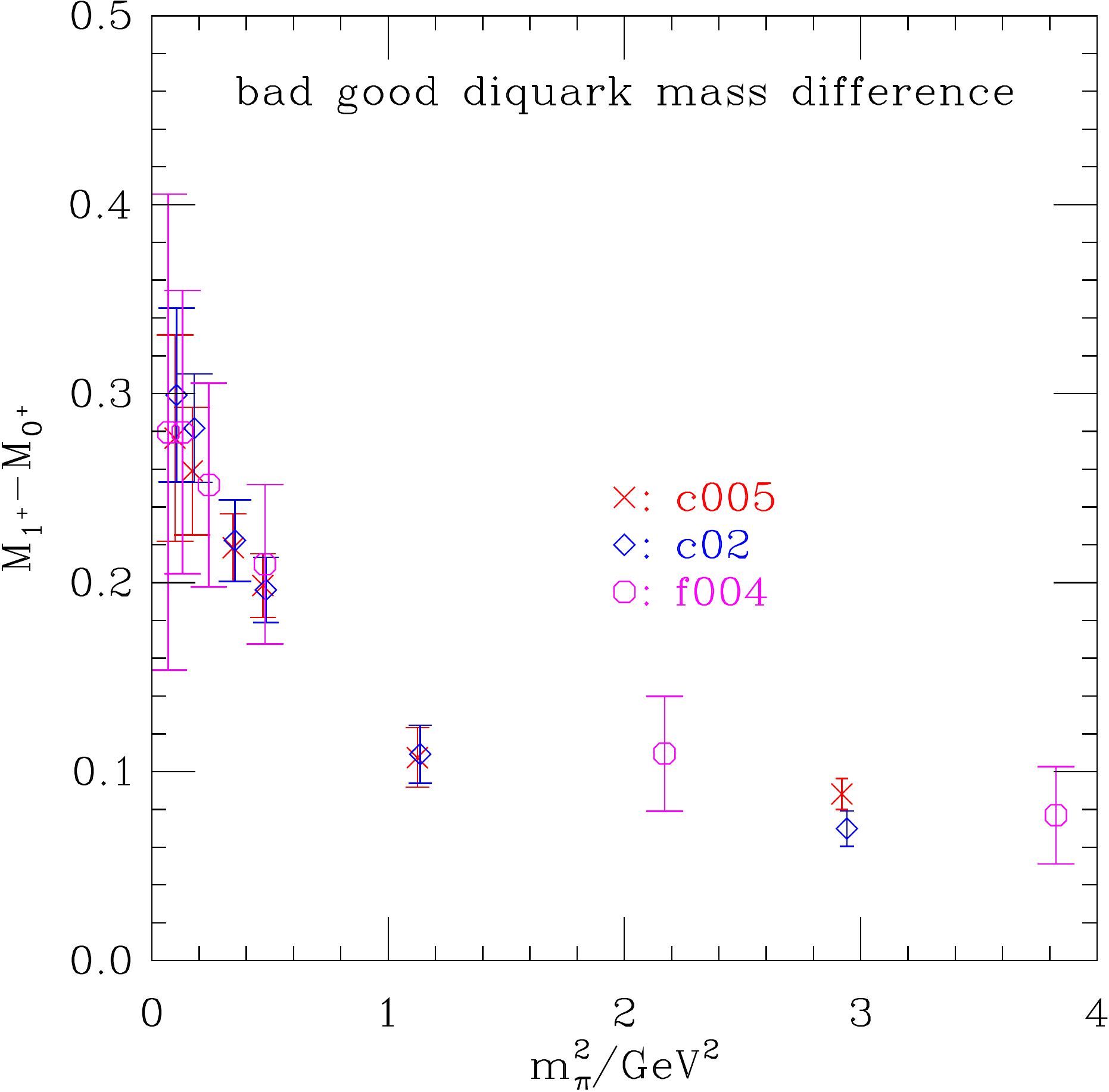}
\end{center}
\caption{Left: The mass difference between bad and good diquarks as a function of the valence quark mass on ensemble c005 and c02.
The results from the two ensembles are in agreement, which means the mass difference has small sea quark mass dependence.
Right: The mass difference (GeV) against the pion mass squared on all three ensembles.}
\label{fig:diff_bad_good}
\end{figure}

For the bad diquark from ensemble c02, a linear extrapolation to the chiral limit with the lowest four data points gives 
$aM_{1^+}=0.644(16)$, which is higher than 0.584(21) from ensemble c005. The bad and scalar diquark mass difference is given in the fourth column
of Tab.~\ref{tab:diquark_mass_c02}. Unlike the absolute value of diquark masses, the diquark mass differences on the two ensembles
are in agreement within statistical uncertainties. In the left graph of Fig.~\ref{fig:diff_bad_good}, the mass differences are plotted against the 
valence quark mass from ensembles c005 and c02. It does not show apparent sea quark mass dependence.

On ensemble f004, the bad diquark mass in the chiral limit is $aM_{1^+}=0.425(24)$ or $M_{1^+}=990(60)$ MeV, which is in agreement with
the result 1022(44) MeV from c005. 
Therefore we do not see discretization effects with our current statistical uncertainties.
The fourth column in Tab.~\ref{tab:diquark_mass_f004} is the bad and good diquark mass difference on the fine lattice.

We plot the bad and good diquark mass difference in physical units against the pion mass squared on all three ensembles 
in the right graph of Fig.~\ref{fig:diff_bad_good}.
All lattice results seem to lie on a universal curve, which means sea quark mass dependence and discretization effects are
small compared with the statistical errors. By using the lowest order relation $m_\pi^2\propto m_q$, Eq.(\ref{eq:ans1}) can be
written as
\begin{equation}
\Delta m=\frac{b_1}{1+b_2 m_\pi^4}.
\end{equation}
This ansatz can fit the data points at $m_\pi^2<1.2(0.6)$ GeV$^2$ and gives 
$M_{1^+}-M_{0^+}=264(14)(285(20))$ MeV ($\chi^2/$dof=3.2/12(1.2/10)).
Alternatively, a linear extrapolation in $m_\pi^2$ with the data points 
at $m_\pi^2<1.2(0.6)$ GeV$^2$ gives $M_{1^+}-M_{0^+}=280(12)(309(25))$ MeV.
The average of the four center values is 285 MeV. Taking the largest statistical error and the largest change
in the center value as the systematic error, we get $M_{1^+}-M_{0^+}=285(25)(45)$ MeV.
This number is a little bigger than the estimation $\sim210$ MeV
in Ref.~\cite{Jaffe:2004ph}, which used masses of baryons with a strange or charm quark.

\begin{figure}
\begin{center}
\includegraphics[height=2.5in,width=0.49\textwidth]{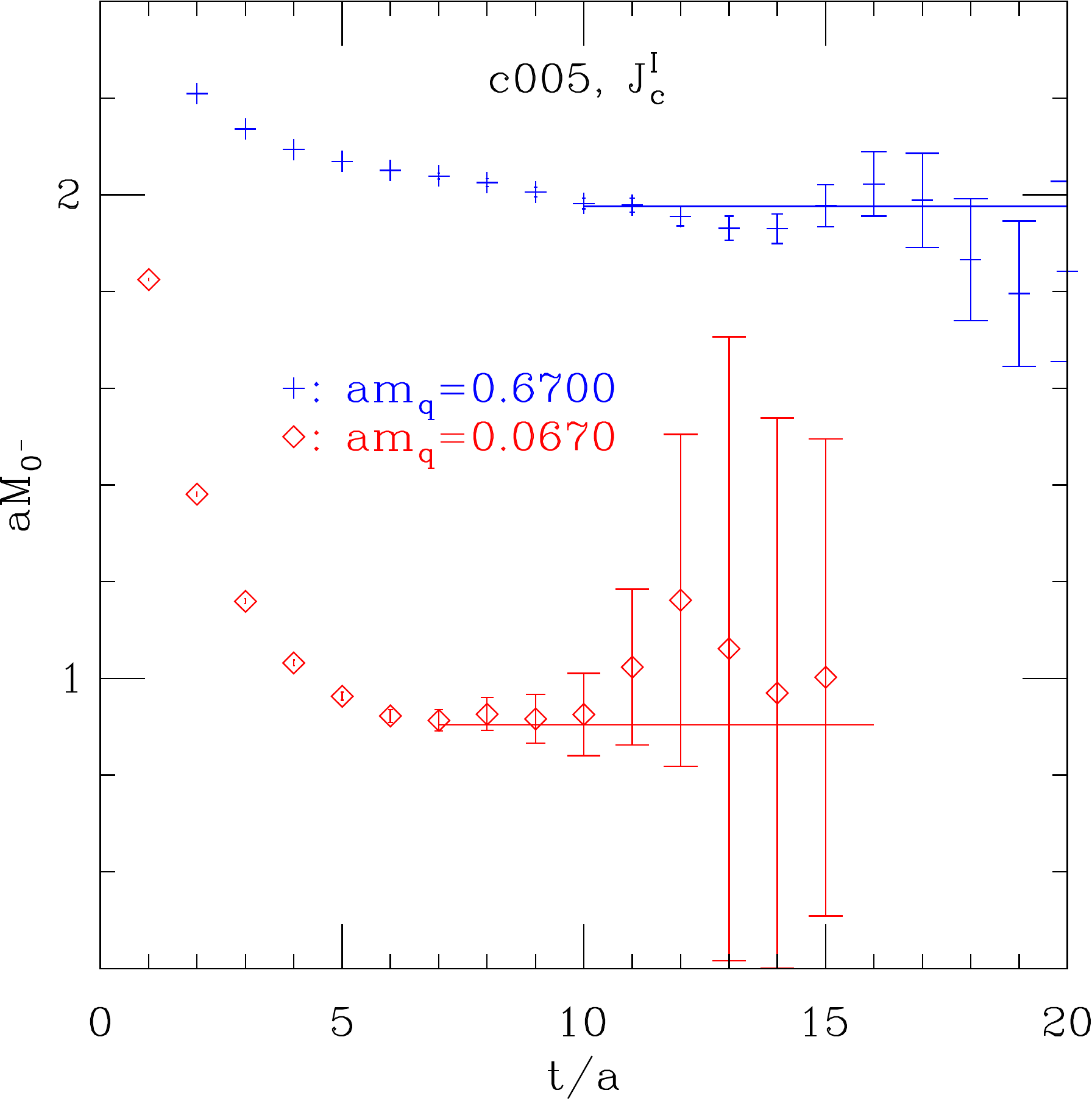}
\includegraphics[height=2.5in,width=0.49\textwidth]{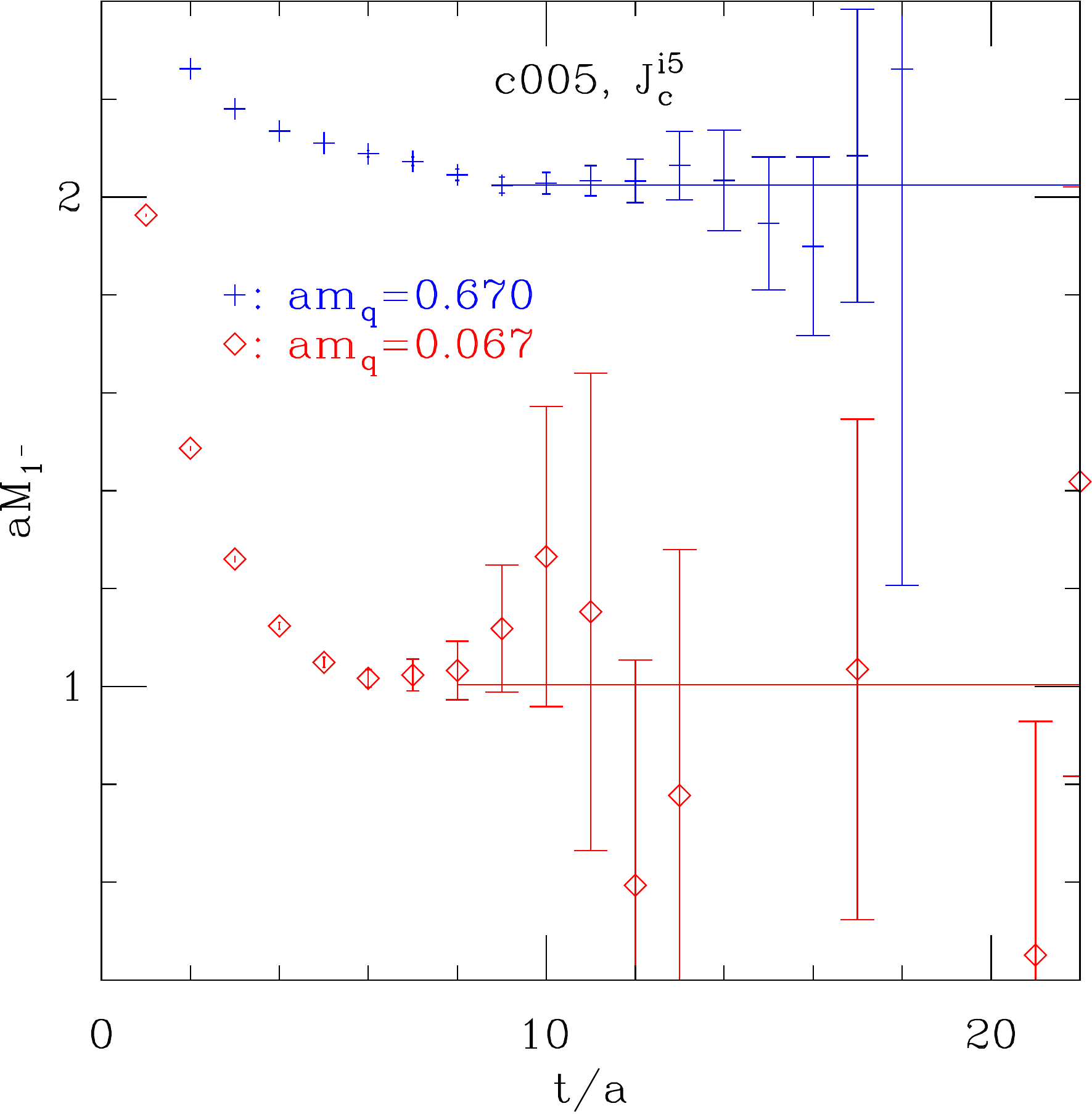}
\end{center}
\caption{Effective masses of the $0^-$ (left) and $1^-$ (right) diquarks at two valence quark masses on ensemble c005. 
The $0^-$ diquark uses correlators from $J^{I}_c$.
The straight lines mark the diquark mass from single exponential fits to the correlation functions. 
The fitting ranges of $t$ are indicated by the length of the lines.}
\label{fig:0minus_c005}
\end{figure}

For the diquark with quantum number $J^P=0^-$, we tried two interpolating operators $J_c^I$ and $J_c^0$. The correlators from both operators
are noisy with the one from $J_c^I$ having a better signal. The masses from single exponential fits to the correlators of $J_c^I$ 
on ensemble c005, c02 and f004
are given in Tabs.~\ref{tab:diquark_mass},~\ref{tab:diquark_mass_c02},~\ref{tab:diquark_mass_f004} respectively.
Examples of the effective mass plateau are shown in the left graph of Fig.~\ref{fig:0minus_c005}.
At some of the small valence quark masses,
no result is obtained due to the bad signal to noise ratio. 
These results confirm that the $0^-$ diquark is heavier than both the $0^+$ and $1^+$ diquarks.

The correlator for the vector diquark is even noisier. 
The extracted diquark masses are listed in Tab.~\ref{tab:diquark_mass} and Tab.~\ref{tab:diquark_mass_c02}
for the two ensembles on the coarse lattice respectively. The right graph in Fig.~\ref{fig:0minus_c005} shows
two examples of the effective mass for this channel.
The vector diquark seems to be heavier than the $0^-$ diquark. But with our statistical uncertainty, it is hard to determine.
On ensemble f004 (see Tab.~\ref{tab:diquark_mass_f004}), the $0^-$ diquark seems to be heavier than the vector diquark.
More statistics are needed to improve the mass plateaus for the $0^-$ and $1^-$ diquarks.

\subsubsection{Diquarks with a strange and a light quark}
Now we turn to diquarks composed of a strange and a light quark. We set $q_1=u$ and $q_2=s$ in the currents given in Tab.~\ref{tab:currents}.
Therefore we use $am_{q_2}=0.0670$ and $am_{q_1}=0.0135$, $0.0243$, $0.0489$, $0.0670$ on the coarse lattice. 
The scalar and bad diquark masses together with their difference
are given in Tabs.~\ref{tab:diquark_us},~\ref{tab:diquark_us_c02} as we vary the mass of $q_1$ on the two ensembles c005 and c02.
\begin{table}
\begin{center}
\caption{Diquark masses and mass difference for various light quark masses on ensemble c005. 
The diquarks are composed of a strange and a light quark.
The first line is a linear extrapolation in $am_{q_1}$ to the chiral
limit.}
\begin{tabular}{cccc}
\hline\hline
$am_{q_1}$       & $aM_{0^+}$($J^{05}_c$) & $aM_{1^+}$($J^i_c$) & $a(M_{1^+}-M_{0^+})$\\
\hline
0.0      & 0.5177(43)  &  0.609(16) &  - \\
0.01350   & 0.5376(42) & 0.633(18) & 0.095(18) \\
0.02430   & 0.5534(38)  & 0.656(14) &  0.103(15) \\
0.04890   & 0.5884(32)  & 0.691(10) &  0.103(11) \\
0.06700   & 0.6166(48) & 0.7300(85) & 0.1134(93) \\
\hline\hline
\end{tabular}
\label{tab:diquark_us}
\end{center}
\end{table}
\begin{table}
\begin{center}
\caption{Diquark masses and mass difference for various light quark masses on ensemble c02. 
The diquarks are composed of a strange and a light quark.
The first line is a linear extrapolation in $am_{q_1}$ to the chiral
limit.}
\begin{tabular}{cccc}
\hline\hline
$am_{q_1}$       & $aM_{0^+}$($J^{05}_c$) & $aM_{1^+}$($J^i_c$) & $a(M_{1^+}-M_{0^+})$\\
\hline
0.0      &  0.579(13) & 0.6883(86)  & -  \\
0.01350   & 0.593(14)  & 0.7037(89) & 0.111(17) \\
0.02430   & 0.602(12)  & 0.7119(72) &  0.110(14) \\
0.04890   & 0.629(10)  & 0.7337(60) &  0.105(12) \\
0.06700   & 0.6453(63) & 0.7574(85) & 0.1121(95) \\
\hline\hline
\end{tabular}
\label{tab:diquark_us_c02}
\end{center}
\end{table}

In the chiral limit of the up quark, we obtain $aM_{0^+}=0.5177(43)$ by doing a linear extrapolation in $am_{q_1}$ on ensemble c005. 
In physical units, it is $906(22)$ MeV. 
For the scalar diquark and strange quark mass difference, one gets $0.5177(43)-0.3351(45)=0.1826(62)$ in lattice units 
or 320(13) MeV, which is of the same size as 312(23) MeV for the scalar diquark composed of two light quarks.
On ensemble c02, this difference is $0.579(13)-0.3443(37)=0.235(14)$ or 411(26) MeV. It is heavier than the result on c005,
showing some sea quark mass dependence. This dependence mainly comes from the scalar diquark mass since the strange quark mass $M_s$ is
not so sensitive to the light sea quark mass (see the end of Sec.~\ref{sec:quark_mass}).

On the fine lattice, we set $am_{q_2}=0.04700$ and $am_{q_1}=0.00677$, $0.01290$, $0.02400$, $0.04700$.
In Tab.~\ref{tab:diquark_us_f004}, we give the diquark masses from ensemble f004.
\begin{table}
\begin{center}
\caption{Diquark masses and mass difference for various light quark masses on ensemble f004. 
The diquarks are composed of a strange and a light quark.
The first line is a linear extrapolation in $am_{q_1}$ to the chiral
limit.}
\begin{tabular}{cccc}
\hline\hline
$am_{q_1}$       & $aM_{0^+}$($J^{05}_c$) & $aM_{1^+}$($J^i_c$) & $a(M_{1^+}-M_{0^+})$\\
\hline
0.0      &  0.3862(93) & 0.453(14)  & -  \\
0.00677   & 0.395(11) & 0.466(16) & 0.071(19) \\
0.01290   & 0.406(10)  & 0.481(15) &  0.075(17) \\
0.02400   & 0.4246(97)  & 0.497(13) &  0.072(17) \\
0.04700   & 0.457(10) & 0.547(15) & 0.090(18) \\
\hline\hline
\end{tabular}
\label{tab:diquark_us_f004}
\end{center}
\end{table}
The scalar diquark and strange quark mass difference is $0.3862(93)-0.2466(81)=0.140(12)$ or 326(29) MeV. 
It agrees with the result 320(13) MeV from c005,
indicating small discretization effect in this difference. Averaging the results from c005 and f004 weighted by their
inverse squared error, one gets $M_{0^+}-M_s=321(12)$ MeV for diquarks composed of a light and a strange quark.

Using the results 320 MeV and 411 MeV from c005 and c02 respectively, we can do a linear extrapolation to 
the light sea quark massless limit: $a(m_{sea}+m_{res})=0$.
What we get is 271(49) MeV. Taking the difference between 320 MeV and 271 MeV as a systematic error, we find
$M_{0^+}-M_s=321(12)(49)$ MeV.
This number is smaller than the estimation $\sim500$ MeV in Ref.~\cite{Jaffe:2004ph}.

For the bad diquark mass, we also do a linear extrapolation in the light valence quark mass and find $0.609(16)$ in the chiral limit
(1.066(37) MeV in physical units) on c005.
Using this number and the chiral limit value of the good diquark mass $0.5177(43)$, one gets the mass difference in the chiral limit
as $0.091(17)$ or $159(30)$ MeV. 
It agrees with the estimation of this diquark mass difference 
(152 MeV) in Ref.~\cite{Jaffe:2004ph} 
obtained from baryon masses in the charm sector.
Compared with the case for diquarks composed of two light quarks, 
this difference decreases as one of the light quark is changed to a strange quark.

On ensemble c02,
the absolute values of the bad and good diquark masses seem heavier than their counterparts on ensemble c005, indicating some
sea quark mass dependence. This is similar to the case for diquarks composed of two light quarks.
The mass difference between the bad and good diquarks on ensemble c02 agrees with that on ensemble c005,
showing that the sea quark mass dependence is smaller in the difference than in the absolute diquark masses.

On ensemble f004, the scalar and bad diquark masses in the chiral limit of $m_{q_1}$ are 900(29) MeV and 1.055(40) MeV respectively.
Both are in agreement with their counterparts from c005. Thus discretization effects are again shown to be small.

\begin{figure}
\begin{center}
\includegraphics[height=2.5in,width=0.49\textwidth]{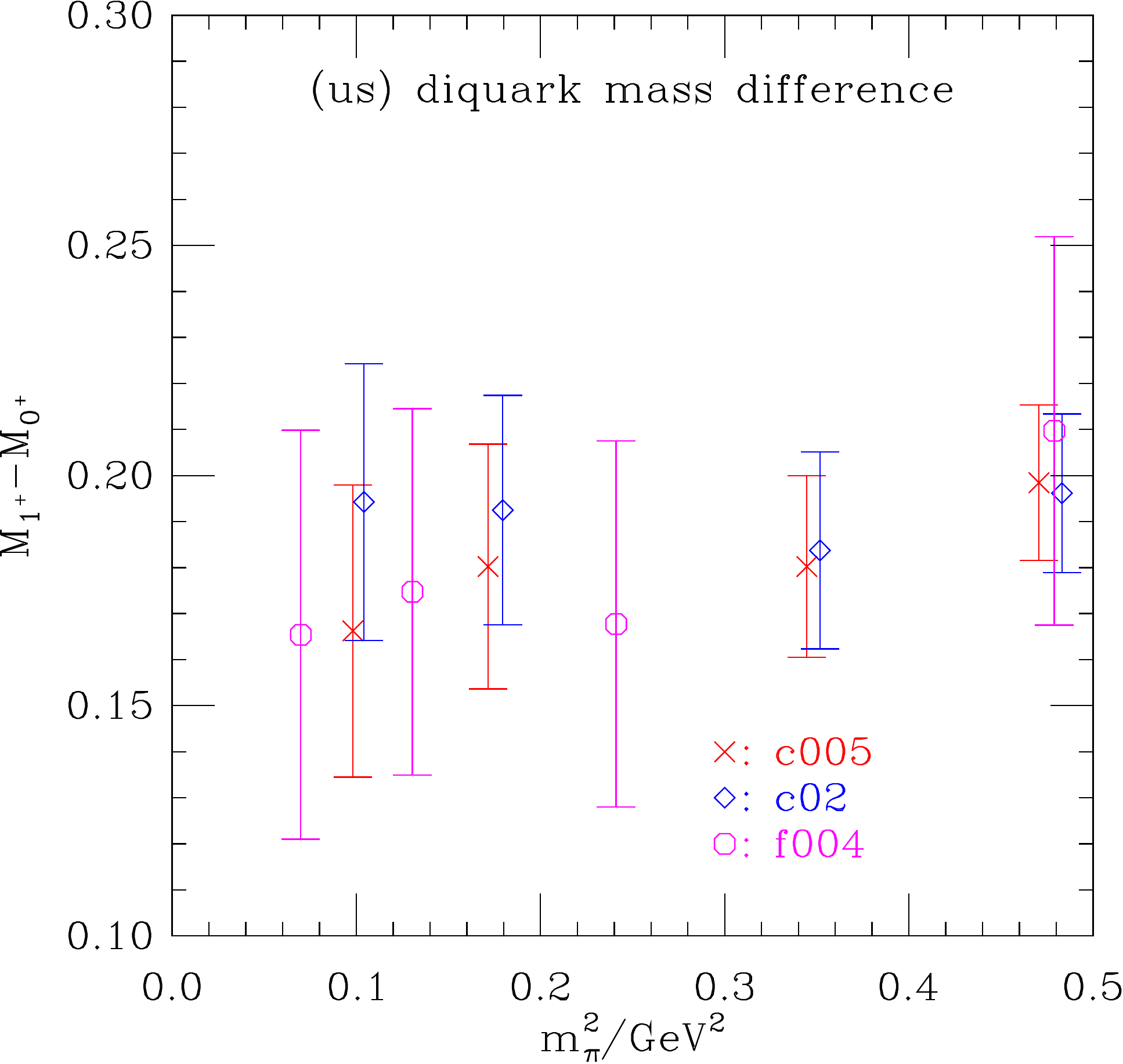}
\end{center}
\caption{The scalar and bad diquark mass difference (GeV) against the pion mass squared on all three ensembles.
The diquarks are composed of a strange and a light quark.}
\label{fig:diff_us_bad_good}
\end{figure}

The scalar and bad diquark mass difference from all three ensembles are plotted in Fig.~\ref{fig:diff_us_bad_good} as a function of 
the squared mass of the pion composed of light quark $q_1$ (see Tabs.~\ref{tab:hadrons},~\ref{tab:hadrons_c02},~\ref{tab:hadrons_f004}).
As we can see from the graph, the dependence of the mass difference on the light quark mass (or pion mass) seems small
with our current statistical uncertainty.
One can fit the difference either with a straight line in $m_\pi^2$ or 
with a constant. The straight line fit gives $M_{1^+}-M_{0^+}=171(18)$ MeV. The constant fit gives $M_{1^+}-M_{0^+}=188(7)$ MeV.
Averaging the two, we get 180(18)(17) MeV. Here the first uncertainty is the bigger one of the two statistical errors 
and the second one is the systematic error from the change in the center values.

\section{Summary}
\label{sec:summary}
Using overlap valence quark on configurations with 2+1 flavors of domain wall sea quarks, 
we calculated the mass and mass difference of various diquarks in the Landau gauge. Also the diquark quark mass difference is computed.
We extrapolate the results to the valence quark chiral limit and check the sea quark mass dependence using two ensembles with a
same coarse lattice spacing.
Discretization effects are examined by working on a fine lattice.

The scalar diquark has the lowest mass which means it is the channel with the strongest correlation. 
The mass difference between the axial vector and scalar diquark (composed of two light quarks)
decreases as the valence quark mass increases.
This was also observed in previous lattice calculations~\cite{Hess:1998sd,Alexandrou:2006cq,Babich:2007ah}.

We see sea quark mass dependence in the absolute values of diquark and quark masses. 
Their masses decrease as the sea quark mass decreases.
This dependence is small (smaller than our statistical error) in diquark mass difference and diquark quark mass difference.
For the diquark composed of a strange and a light quark, the mass difference between the scalar diquark and the strange quark shows
some sea quark mass dependence. From our data we do not expect this difference to increase as the light sea quark mass 
lowers to the physical value.
Within our limited statistics on the fine lattice, we do not see apparent discretization effects in all our results.

Our final results of the mass differences are given in Tab.~\ref{tab:summary}.
\begin{table}
\begin{center}
\caption{Diquark mass difference and diquark quark mass difference (MeV).
The diquarks are either composed of two light quarks or composed of a strange and a light quark.
The first (or the only) error is statistical and the second (when there is one) is a systematic error.}
\begin{tabular}{cccc}
\hline\hline
  (ud)               & $M_{0^+}-M_{q}$  & $M_{1^+}-M_{0^+}$ \\
 this work           &  304(17)         & 285(25)(45)       \\
 \cite{Jaffe:2004ph} &  $\sim$310       & $\sim$210         \\
\hline
   (us)              & $M_{0^+}-M_{s}$  & $M_{1^+}-M_{0^+}$ \\
 this work           &  321(12)(49)     & 180(18)(17)       \\
 \cite{Jaffe:2004ph} &  $\sim500$       & $\sim$150         \\
\hline\hline
\end{tabular}
\label{tab:summary}
\end{center}
\end{table}
In the chiral limit of the valence quark mass,
We find the diquark mass difference $M_{1^+}-M_{0^+}=285(25)(45)$ MeV and diquark quark mass difference $M_{0^+}-M_q=304(17)$ MeV
for diquarks composed of two light quarks.
For diquarks composed of a strange and a light quark, we obtain 
$M_{1^+}-M_{0^+}=180(18)(17)$ MeV and $M_{0^+}-M_s=321(12)(49)$ MeV. 
Here when there are two uncertainties, the first one is statistical and the second one is a systematic error 
estimated from different extrapolations to the chiral limit or from light sea quark mass dependence.
In general, the results of these mass differences agree with the estimations from hadron spectroscopy in Ref.~\cite{Jaffe:2004ph}.
The exception is $M_{0^+}-M_s=321(12)(49)$ MeV for the scalar diquark composed of a strange and an up quark, 
which is smaller than the estimation $\sim500$ MeV in Ref.~\cite{Jaffe:2004ph}.

To better control the light sea quark mass dependence and finite lattice spacing effects in our work, calculations at
another sea quark mass are needed, and more statistics on the fine lattice should be added. 
It might be interesting to calculate these differences in other gauges to check the gauge dependence.

\section*{Acknowledgements}
We thank RBC-UKQCD collaboration for sharing the domain wall fermion configurations. 
This work is partially supported
by the National Science Foundation of China (NSFC) under Grants 11575197, 10835002,
11405178 and 11335001 and by joint funds of NSFC under contracts No. U1232109. 
MG and ZL are partially supported by the Youth Innovation Promotion Association of CAS (2015013,2011013).
YC and ZL acknowledge the support of NSFC and DFG (CRC110).
Parts of the fittings were done by using Meinel's public code~\cite{QMBF}.

\end{document}